\documentstyle[pre,aps,epsfig,multicol]{revtex} 

\begin{document}
\draft \title{Liquid crystal-solid interface structure at the
  antiferroelectric-ferroelectric phase transition} \author{D. Coleman, S.
  Bardon, L. Radzihovsky, G. Danner, and N. A. Clark} \address{Condensed
  Matter Laboratory, Department of Physics,
  University of Colorado,\\
  Boulder, CO 80309 USA} \date{30 December 2002} \maketitle

\begin{abstract}
  Total Internal Reflection (TIR) is used to probe the molecular organization
  at the surface of a tilted chiral smectic liquid crystal at temperatures in
  the vicinity of the bulk antiferroelectric-ferroelectric phase transition.
  Data are interpreted using an exact analytical solution of a real model for
  ferroelectric order at the surface.  In the mixture {\bf T3}, ferroelectric
  surface order is expelled with the bulk ferroelectric-antiferroelectric
  transition.  The conditions for ferroelectric order at the surface of an
  antiferroelectric bulk are presented.
\end{abstract}


\begin{multicols}{2}
\narrowtext

Antiferroelectric and antiferromagnetic ordering occur when there is a
sufficiently strong local interaction favoring antiparallel neighboring
dipoles or spins.  The result is macroscopic ordering that is nonpolar in the
absence of external fields.  By contrast, material surfaces are inherently
polar, there being an obvious direction from one material to another.
Therefore at the surface of antiferroelectrics, there is an intrinsic
competition between the bulk antiferroelectricity and the surface
ferroelectric ordering tendency.  Liquid crystals (LCs) are an attractive
system for studying this behavior because their large optical anisotropy and
sensitivity to surface forces make effective study of surface states possible.
In this paper, we investigate surface structure of tilted chiral smectic LCs,
two-dimensional polar fluid layers of rod-shaped molecules that order into
either ferroelectric or antiferroelectric bulk structures.

\begin{figure}[htbp]
\epsfig{file=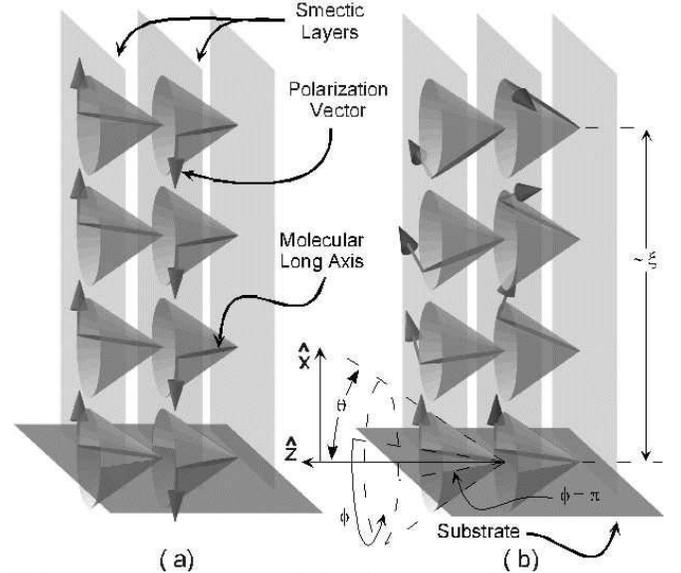,width=\linewidth}
\caption{(a) Schematic of the director and polarization profiles for the 
  uniform anticlinic tilted chiral smectic with boundary surfaces
  perpendicular to the smectic layers. We assume that the dominant surface
  interaction energy is the planar interaction favoring molecular alignment
  along the surface.  (b) The model for a synclinic distortion at the surface
  of an anticlinic bulk smectic material.  The polar nature of the surface
  favors a certain sign of polarization in competition with the bulk
  antiferroelectricity. In the model for the distorted state, the polarization
  makes an angle $\phi$ with the surface normal and the distortion has a
  penetration depth $\xi$.}
\label{Model}
\end{figure}

In a given tilted smectic liquid-crystal layer, $\hat{n}$, the orientation of
the mean long molecular axis is given by $\theta$, the fixed angle of tilt
relative to the layer normal $\hat{z}$, as shown in Fig.~\ref{Model},
and by $\phi(x)$, the azimuthal orientation about $\hat{z}$, a Goldstone
variable degenerate in the free energy in an infinite sample.  Chiral tilted
smectic layers are ferroelectric, the lack of mirror symmetry allowing within
each layer a spontaneous polarization, ${\bf P}$, mutually perpendicular to
$\hat{n}$ and $\hat{z}$ (see Fig.~\ref{Model}). This enables coupling of $\phi$
to electric field {\bf E} applied in the plane of the layer, tending to
minimize ${\bf P \cdot E}$. In the synclinic ferroelectric (SYN, Sm-{\em C}\,*)
tilted chiral smectic phase\cite{meyer}, adjacent layers are tilted in the
same direction ($\Delta \phi = 0$), whereas, in the anticlinic antiferroelectric
(ANTI, Sm-{\em C}$_A$\,*) phase\cite{fukuda1}, adjacent layers are tilted in
opposite directions ($\Delta \phi = \pi$).

We consider here LC-solid surface systems wherein the molecules at the surface
prefer (planar) alignment, i.e., with $\hat{n}$ parallel to the substrate
plane. In a Sm-{\em C}\,* with the layers oriented perpendicular to the
substrates (bookshelf geometry) and with the helix unwound (i.e., surface
stabilized\cite{surfacestabilized}), such planar alignment will generate {\it
  two} preferred $\hat{n}$ orientations corresponding to the intersection of
the tilt cone with the substrate [$\phi(0) = 0$ and $\phi(0) = \pi$ in
Fig.~\ref{Model}]. Additionally, LC chirality couples with the polar nature of
the surface to produce a surface energy difference $\Delta U_{\rm Surf} =
U_{\rm Surf}(0)-U_{\rm Surf}(\pi) $ between these two states, tending to
produce a uniform synclinic order at the surface. By contrast, in the Sm-{\em
  C}$_A$\,* bookshelf geometry, the bulk anticlinic ordering is basically
incompatible with the synclinic order preferred by the
\noindent
surface.  A strong polar surface interaction will cause the system to minimize
its energy by adopting synclinic order near the surface, lowering the surface
energy at the expense of increasing that of the anticlinic bulk
[Fig.~\ref{Model}(b)].  With sufficiently small $\Delta U_{\rm Surf}$, the
ferroelectric order at the surface may be expelled by the antiferroelectric
bulk.  The geometry will then be the configuration of Fig.~\ref{Model}(a) in
which the LC is uniformly anticlinic with $\phi(0)$ equal to $0$ and $\pi$ in
adjacent layers, and having planar alignment of $\hat{n}$.  We have carried
out total internal reflection (TIR) measurement of the optic axis director
orientation in the LC-solid interface in bookshelf tilted chiral smectic cells
with planar aligned surfaces\cite{Xue,Zhiming}. This technique enables us to
readily distinguish synclinic surface and anticlinic surface states. The
results are interpreted using an exact analytic solution of a realistic
theoretical model of surface ordering.

\begin{figure}[tbp]
\centering
    \epsfysize = 120mm
    \epsfbox{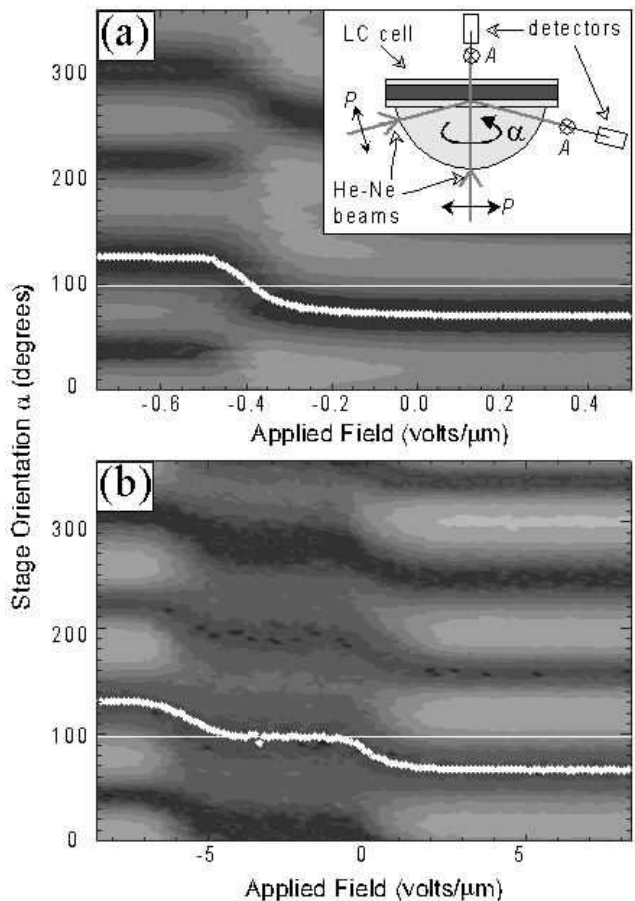}
\caption{
  Contour plot of the depolarized intensity of the TIR signal vs applied
  voltage and stage angle in the SYN (a) and ANTI (b) phases as a function of
  voltage.  The layer normal is indicated by the horizontal solid line and the
  orientation of the optical axis, $\beta$, is indicated by the open diamonds.
  Inset: Setup for the TIR study of the LC orientation near the glass surface
  via depolarization of the reflected He-Ne beam. The beam incident along the
  hemisphere axis probes the orientation of the optical axis of the bulk LC.
  }
\label{contourplots}
\end{figure}

The TIR setup is presented in the inset to Fig.~\ref{contourplots}.  A
\noindent
He-Ne laser beam polarized in the plane of incidence illuminates the LC cell
through a high refractive index hemisphere ($n=1.79$) at a fixed angle of
incidence of $75^{\circ }$. The light is totally reflected at the glass/LC
interface, and an evanescent wave (probe beam) penetrates a depth $\lambda \sim
1000\;{\rm {\AA}}$ into the LC.  Hereafter, we will denote the depolarization
ratio $R(\alpha )$ as the ratio of the detected {\em s}-wave (out of plane)
intensity to the incident {\em p}-wave (in plane) intensity. $ R(\alpha )$,
measured as a function of $\alpha,$ the angle between $\hat{z}$ and the TIR plane
of incidence, gives information about the orientation of the LC close to the
TIR interface. Minima in $R(\alpha )$ occur when an optical axis is rotated to be
either parallel or perpendicular to the TIR plane of incidence.  That is, a
uniform director along $\hat{z}$ produces four zero-intensity minima in
$R(\alpha)$ at $\alpha=0,\pi/2,\pi,3\pi/2$.  As the director tilts out of the
$\hat{z}-\hat{y}$ plane by an angle $\gamma$ [for a Sm-{\em C}, $\tan
\gamma=\sin\theta\sin\phi (1-\sin^2\theta\sin^2\phi)^{-1/2}$] the director is no longer
perpendicular to the plane of incidence at $\alpha=\pi/2,3\pi/2$.
$R(\alpha=\pi/2,3\pi/2)$ becomes nonzero and grows rapidly with increasing $\gamma$,
evolving continuously to become a global maxima for $\gamma>17^\circ$ (for a material
with average index of refraction, $n_{avg}$, equal to $1.58$ and
birefringence, $\Delta n$, equal to $0.1$).  We determined the orientation of the
bulk optical axis of the LC by using an additional laser beam at normal
incidence to the cell plates to probe the cell birefringence via measurement
of the transmission [$T(\alpha)$] between crossed polarizer and analyzer.
$T(\alpha)=0$ corresponds to having a uniformly oriented optical axis that
projects onto the {\it y}-{\it z} plane parallel to the polarizer or analyzer.

We performed experiments in the (SYN, Sm-{\em C}\,*) and (ANTI, Sm-{\em
  C}$_A$\,*) phases of the three component mixture, {\bf T3}
\cite{fukAsiaDisp,fukHerm}, which has the phase sequence: Iso
$\stackrel{69^\circ{\rm C}} {\longleftrightarrow}$ Sm-{\em A}\,*
$\stackrel{64^\circ{\rm C}} {\longleftrightarrow}$ Sm-{\em C}\,*
$\stackrel{43^\circ{\rm C}} {\longleftrightarrow}$ Sm-{\em C}$_A$\,*. The
Sm-{\em C}\,* phase can be expected to have SYN order at the surface and a
phase transition to a bulk ANTI phase that may or may not be accompanied by a
change in the surface ordering. The high index glass substrate at the TIR
interface was coated with a $590-{\rm {\AA}}$-thick transparent conducting
indium tin oxide (ITO) layer ($n=1.96$) and then a $150-{\rm {\AA}}$-thick
rubbed nylon (Du Pont Elvamide 8023) alignment layer to produce uniform
smectic layering.  The glass/ITO/nylon/LC assembly produces a TIR condition
only at the ITO/nylon interface.  The other cell surface was coated with
unrubbed nylon. The bulk optic axis reorientation saturates with $\hat{n}$ on
opposite sides of the tilt cone, $\phi = (0,\pi)$ at $\pm 8\;{\rm V}/\mu{\rm
  m}$, respectively, enabling a determination of the layer normal from the
average optic axis orientation and of the tilt angle $\theta$, half the
difference in the optic axis orientations. For {\bf T3}, $\theta = 31^\circ$
in the Sm-{\em C}\,* ($T = 48^\circ{\rm C})$ and $\theta = 36^\circ$ in the
Sm-{\em C}$_A$\,* ($T = 25^\circ {\rm C}$).  The {\bf T3} cells are surface
stabilized.

Figure \ref{contourplots} shows contour plots of $R(\alpha)$ vs $\alpha$ and
applied (decreasing) electric field ${\bf E}$ at $48 ^\circ {\rm C}$ and $25
^\circ {\rm C}$, well into the Sm-{\em C}\,* and Sm-{\em C}$_A$\,* phase
ranges, respectively. In both phases there are four minima in R($\alpha$) at
zero volts and at the positive and negative saturated voltages indicating that
the average optic axis within $1000\;{\rm {\AA}}$ of the surface is uniform
and parallel to the surface. In the Sm-{\em C}\,* phase at ${\bf E}={\bf 0}$
the optic axis at the surface is rotated by $27^\circ$ from the layer normal,
close to $\phi = 0$ on the tilt cone and parallel to the surface. Application
of ${\bf E} > {\bf 0}$ produces little change, but application of ${\bf E} <
{\bf 0}$ above a threshold magnitude $|E| \approx 0.4\;{\rm V}/\mu{\rm m}$
favoring $\phi = \pi$ overcomes the polar surface pinning to stabilize $\phi =
\pi$ at the surface and reorient the bulk Sm-{\em C}\,*.  During switching,
the minima corresponding to the optical axis perpendicular to the plane of
incidence disappear, indicating tilt relative to the surface as the molecules
reorient on the smectic-{\em C} cone. By contrast, in the Sm-{\em C}$_A$\,*
phase, the surface optic axis is nearly along the bulk layer normal at ${\bf
  E}={\bf 0}$ and the maxima in $R(\alpha)$ are reduced because of the smaller
birefringence of the antiferroelectric (AF) ordering at the surface.
Application of either sign of ${\bf E}$ produces a thresholded transition to
ferroelectric (FE) order at the surface that is observed simultaneously in the
bulk Sm-{\em C}$_A$\,*.  Furthermore, all four minima of $R(\alpha)$ maintain
roughly equal intensity throughout the switching indicating that there is
never any tilt of the optic axis at the surface in the AF phase.

\begin{figure}[tbp]
\centering
    \epsfysize = 60mm
    \epsfbox{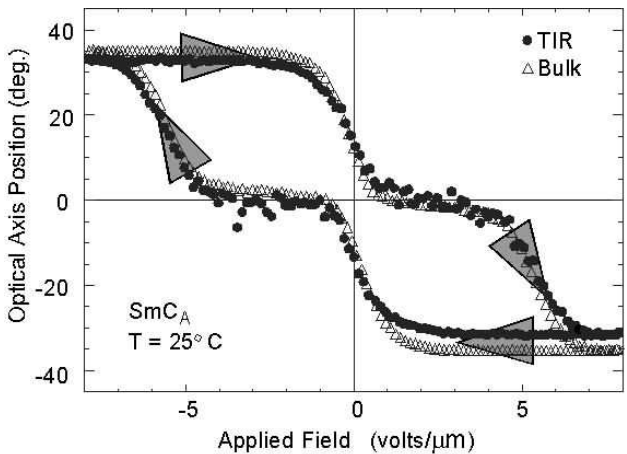}
\caption{
  Evolution of the optical axis relative to the layer normal ($\beta$) with
  applied field in the Sm-{\em C}$_A$\,* phase. The bulk signal displays the
  expected three-state switching with large hysteresis. The surface
  orientation follows the bulk to within $4^\circ$.}
\label{opticaxis}
\end{figure}

In Fig.~\ref{opticaxis}, we present the evolution of the orientation,
\noindent
 $\beta$, of the optical axis
relative to the layer normal in the surface and bulk at $25^\circ{\rm C}$ in the
Sm-{\em C}$_A$\,* phase for a voltage ramp from $-9$ to $+9\;{\rm V}/\mu{\rm
  m}$ at $0.5\;{\rm mHz}$.  In response to this ramp, the bulk and surface
orientations both exhibit a hysteretic transition between AF and FE states.
For both positive and negative voltages, the surface saturates to a smaller
angle than the bulk ($\bracevert\beta_{\rm surf}\bracevert < \bracevert\beta_{\rm
  bulk}\bracevert$). In the surface-preferred positive voltage state, the
surface saturates to a tilt angle of $32^\circ$, $4^\circ$ less than the bulk.
However, in the surface disfavored negative voltage state, the surface
saturates to a tilt angle of $34^\circ$, only $2^\circ$ less than the bulk. Upon
${\bf E} \to {\bf 0}$, the surface optical axis orientation does not return
completely to be along the layer normal, indicating some kind of remnant FE
order characteristic of the state that it was previously in. The small remnant
angle ($\sim 2^\circ$) could be accounted for by a weakly polar state having ${\bf
  P}$ at the surface rotated a few degrees from the ANTI state.  A $40-{\rm
  {\AA}}$-thick strongly FE remnant layer (i.e., with $\phi = \pi$) could also
produce such a rotation.  $R(\alpha)$ data such as in Fig.~\ref{contourplots} were
measured vs temperature near $T=T_{\rm SA}$ (the ANTI to SYN phase transition
temperature), where, because of the ambivalence of the bulk ordering, one
might expect SYN surfaces with ANTI bulk. However, measured bulk and surface
optic axis orientations show a similar temperature dependence and indicate
that if the state with SYN surfaces and ANTI bulk occurs, it is only over a
small temperature range near $T_{\rm SA}$ ($T_{\rm SA} -T < 1.8 ^\circ{\rm C}$).

Further interpretation of this temperature behavior requires a calculation of
the structure of $\phi(x)$ using a quantitative model of the bulk and surface
energetics.  We model the $\phi$ dependence of the surface energy per unit area
by
\begin{equation} \label{eq:SurfPot}
U_{\rm Surf} = -S_p \cos{\phi} - S_{np} \cos{2\phi},
\end{equation}
having polar ($S_p$) and nonpolar ($S_{np}$) interaction coefficients.  We
model the $\phi$ dependence of the bulk energy per unit volume by nearest
neighbor layer interactions of the form
\begin{eqnarray} \label{eq:BulkIntLay}
U_{\rm Bulk} = U_p (1+\cos{(\phi_\ell - \phi_{\ell +1})}) 
\nonumber\\
+ U_{np} (1-\cos{2(\phi_\ell - \phi_{\ell +1})})
\end{eqnarray}
with $U_p$ and $U_{np}$ chosen so that $U_{\rm Bulk}$ exhibits a local minimum
for the SYN orientation and a global minimum for the ANTI orientation; $2U_p$
is the energy difference between the ANTI and SYN states in the absence of a
surface, while $U_B = {(U_p+4U_{np})^2}/{8U_{np}}$ is the height of the energy
barrier between these states.

To model the surface order in the ANTI phase, two configurations shown in
Figs.~\ref{Model}(a,b) need to be considered.  The uniform configuration is an
ANTI state right down to the surface that minimizes the bulk energy and the
nonpolar surface energy while the polar surface energy alternates between the
stable and unstable minimum values ($U_{\rm Surf} = -S_{np} -S_{p}$ and
$U_{\rm Surf} = -S_{np} +S_{p}$ in adjacent layers).  The bulk energy of the
uniform configuration is zero, $U_{\rm Bulk} = 0$, with a total energy $E_U =
-L N d S_{np}$ where $L$ is the cell length, $N$ is the number of layers in
the cell, and $d$ is the layer thickness.  Both configurations have an average
optic axis in the $\hat{z}-\hat{y}$ plane, thus satisfying the zero-pretilt
required by the four minima in $R(\alpha)$ [Fig. \ref{Model}(b)].

In the distorted configuration, a bulk ANTI state deforms into a SYN state at
the surface, where in general $\phi (0) \neq 0$.  We assume that the $\phi_l(x)$ for
all even layers are identical and the negative of their odd neighbors, i.e.,
$\phi_\ell(x) = - \phi_{\ell+1}(x)$, reducing the problem to the determination of the
single field, $\phi(x)$, degree of freedom. Working in the one-elastic constant
approximation, the bulk free energy arising from elastic deformation,
layer-layer interaction, and polarization self interaction is given by
\begin{eqnarray} \label{eq:FEBulk}
H_b[\phi(x)] =
        A \int_{0}^{\infty} dx \lgroup
        {\frac{1}{2}}K(\frac{d \phi}{dx})^2 + \nonumber\\
        U_{\rm Bulk}(\phi)+
        \frac{1}{2 \epsilon_o}{\mathcal P}^2 \cos^2(\phi)\rgroup,
\end{eqnarray}
where $A = L N d$.  The form of the polarization self-interaction energy (the
third term)\cite{Zhiming2} is the same as the polar term in the interlayer
potential so that $U_{p} \to U_{p} + {{\mathcal P}^2}/{4\epsilon_o}$ in the
following discussion. A solution, $\phi(x)$, minimizing the bulk energy
$H_b[\phi(x)]$, subject to the constraint at the surface, $\phi(0)$, has a
sigmoidal, soliton-like form characterized by a penetration length $\xi \approx
\sqrt{{K}/{U_p +4 U_{np}}}$ and can be found exactly via standard methods.
Inserting this solution $\phi(x)$ into eq.~\ref{eq:FEBulk}, we find the bulk
energy as a function of the surface polarization boundary condition, $\phi(0)$,
\begin{eqnarray} \label{eq:HBulkSol}
& & H_b[\phi(0)] = \hspace{0.5cm} {A \frac{\sqrt{K}}{2}
\tan{\phi(0)}}\nonumber\\ & & \times { \lgroup
U_p+U_{np}+U_p\cos{2\phi(0)}-U_{np}
            \cos{4\phi(0)}
            \rgroup ^{\frac{1}{2}}}\nonumber\\
& & - A \frac{U_p}{4}  \sqrt{\frac{2 K}{U_{np}}} \arctan\!\!
        {\left( \!
            \frac{2\sqrt{\frac{U_{np}}{U_p}}\sin{\phi(0)}}
            {\sqrt{U_p+2 U_{np}-2 U_2\cos{2\phi(0)}}}
\right)}
\end{eqnarray}
Combining this bulk energy with the energy of the surface
(eq.~\ref{eq:SurfPot}), we can exactly find the energy of the distorted state
and the equilibrium surface angle $\phi(0)$ by minimizing the total energy $E_D =
H_b (\phi(0)) + A U_{\rm Surf}(\phi(0))$ over $\phi(0)$.

\begin{figure}[htb]
\centering
    \epsfysize = 50mm
    \epsfbox{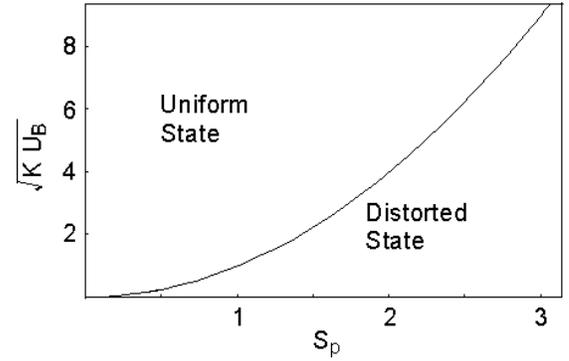}
\caption{ 
  Phase diagram for uniform and distorted states as a function of the polar
  surface potential ($S_p$) and the barrier between the synclinic [Fig.
  ~\ref{Model}(b)] and anticlinic [Fig. ~\ref{Model}(a)] states ($U_B$).}
\label{PhaseDi}
\end{figure}

The bulk Sm-{\em C}\,* to Sm-{\em C}$_A$\,* transition is first order as
evidenced by coexistence of Sm-{\em C}\,* to Sm-{\em C}$_A$\,* domains in
electro-optic experiments.  Consequently, $U_B > 2 U_p$ (where $U_p$ includes
the polarization self interaction) so that $U_B$ always provides the dominant
contribution to the distortion energy. For the range of parameters relevant to
our experiment, it can be shown that $H_b$ is proportional to $\sqrt{K U_B}$.
Because both the uniform and distorted states minimize the planar interaction
energy in Eq.~\ref{eq:SurfPot}, $S_{np}$ gives approximately equal
contribution to the energy of both the uniform and distorted states.  This
leaves only $\sqrt{K U_B}$ and $S_p$ as competing energies to determine the
stability of the distorted state and leads to the phase boundary illustrated
in Fig.~\ref{PhaseDi}. 
There is a critical value, $S_{pc} \propto \sqrt{K U_B}$,
above which the polar surface interaction is strong enough to induce the
ferroelectric distortion. For LC-surface systems where $\sqrt{K U_B}>S_p$, the
surfaces are too weak to pay the energy cost of the ferroelectric distortion
and the LC will be uniformly antiferroelectric with surface optic axis along
the layer normal.

This analysis shows that it is essential to include the nonpolar term in the
bulk free energy in calculations of surface states of AFLCs.  Although with
high ${\bf P}$ there is always a polar term in the free energy,
this contribution will be unimportant in low ${\bf P}$ mixtures
near the Sm-{\em C}\,* to Sm-{\em C}$_A$\,* transition, but the barrier
between the synclinic and anticlinic states (provided by the nonpolar term)
should still be present and will determine the surface state.

One possible explanation for the difference between the bulk and the surface
optic axis orientation at zero applied field is a weakly ferroelectric surface
state characterized by $\phi(x=0) \approx {\pi}/{2}$ and a long penetration depth
into the bulk.  However, our model demonstrates that FE surface states with
large $\phi$ (${\bf P}$ only slightly rotated from the AF orientation)
are energetically unfavorable compared to the uniform state. Consequently, for
a ferroelectric layer to explain a small optical axis rotation, the layer must
have small $\phi(x=0)$ and small decay length. From measurements of the
polarization ($P = 170\;{\rm nC}/{\rm cm}^2$\cite{vshape1}) and the critical
field for inducing the ferroelectric state ($5\;{\rm V}/\mu$), we deduced the
value of parameter $U_1= 8.5\;{\rm kJ}/{\rm m}^3 \sim U_B$. We then used a
typical value for a liquid crystal elastic constant to estimate the decay
length $\xi \approx 1000\;{\rm {\AA}}$ of the ferroelectric surface order in the T3-nylon
system. This decay length is similar to the TIR probe depth so that a TIR
measurement of a T3 cell in the distorted state will detect an optical axis
oriented along the tilt cone.  Consequently, according to this model, the
T3-nylon system at $25 ^\circ{\rm C}$ is in the uniform state.

The discovery of a material with a second-order Sm-{\em C}\,* to Sm-{\em
  C}$_A$\,* phase transition and small polarization, ${P}$, would
present the opportunity to observe the distorted state as described by our
model.  As long as bulk energy dominates, $\sqrt{K U_B}>S_{pc}$ at $T = T_c$,
the cell will remain in the uniform state until the bulk transition is
reached. However, at a second-order phase transition, $U_B$ vanishes as $T\to
T_c^-$. At some finite temperature below the transition, $\sqrt{K U_B}$ would
be less than $S_p$ and the distortion would appear.  Furthermore, as the LC
approaches the second order transition from below, the decay length diverges.
This would have the effect of reducing the AF to FE transition temperature in
thin cells, where the cell thickness is of order of the decay length.

We conclude that in systems with surface interactions strong enough to induce
ferroelectricity, the ferroelectricity will extend into the bulk over a
distance $\xi$ governed by $K$ and $U_B$. For a first-order Sm-{\em C}$_A$\,*
to Sm-{\em C}\,* transitions, the polar surface term must be larger than
$\sqrt{K U_B}$ to induce the ferroelectric state. However, for second-order
transitions, the distorted state always appears below $T_c$.  Consequences of
this are a pretransitional ferroelectric ordering at the surface and the
depression of $T_c$ in thin cells. Based on our measurements, we conclude that
the polar surface interactions are not strong enough to induce ferroelectric
surfaces in the T3-nylon system. It may be possible to increase the polar
surface interaction, perhaps by using a material with a larger surface
electroclinic effect\cite{Xue2,Shao}.

This work has been supported by NSF MRSEC Grant No.\ DMR-9809555 and AFOSR
MURI Grant No.\ F49620-97-1-0014.

\end{multicols}
\end{document}